\documentstyle[epsfig,12pt]{article}

\begin{document}

\begin{center}

{\bf
\title "Energy generation in stars}

\bigskip

\bigskip
\author "B.V.Vasiliev
\bigskip

{Institute in Physical-Technical Problems, 141980, Dubna, Russia}
\bigskip

{vasiliev@dubna.ru}
\end{center}

\bigskip

\begin{abstract}

It is a current opinion that thermonuclear fusion is the main
source of the star activity. It is shown below that this source is
not unique. There is another electrostatic mechanism of the energy
generation which accompanies thermonuclear fusion. Probably, this
approach can solve the solar neutrino problem.

\end{abstract}
\bigskip

The basic source of energy in stars is the nuclear energy. At high
temperature inside a star there are conditions for the
thermonuclear fusion. More heavy nuclei have greater bonding
energy and greater mass defect. The energy related to this mass
defect becomes free at fusion of heavy nuclei and it is used up
for a heating of a star and a radiation. This is generally known.

The thermonuclear fusion can be considered as a process of a
growing of the averaged mass number A and the averaged charge  Z
of nuclei in plasma.

The nuclear mass table data show that the bonding energy which
becomes free at thermonuclear fusion  depend linearly on the
averaged charge of nuclei

\begin{equation}
E_{m}\approx - 3\cdot 10^{-5}Z~erg
\end{equation}

But, at consideration of the thermonuclear fusion as an unique
process of the energy generation in stars, one looses sight of the
fact that there is other additional parallel mechanism of the
energy generation - a decrease in the gravitational energy of a
star, which is caused by of decrease in the electrostatic energy
of plasma and is a result of nuclear charge build-up.

According to the virial theorem \cite{LL8},\cite{BV-usp} the full
energy of plasma is equal to the half of its potential energy. The
potential energy of plasma inside a star
\cite{BV-NC},\cite{BV-book} related to one nucleus is

\begin{equation}
E_z =
\frac{U}{N_{\star}}=-\frac{GM_{\star}^2}{2R_{\star}N_{\star}}
\end{equation}

According to \cite{BV-NC}, the steady-state value of the star mass
is

\begin{eqnarray}
M_{\star}=1.5^6 \sqrt{\frac{10}{\pi^3}} \biggl(\frac{\hbar c}
{Gm_p^2}\biggr)^{3/2} \frac{m_p}{(A/Z)^2}\approx \nonumber\\
\approx 6.47 \biggl(\frac{Z}{A}\biggr)^2 {M_{Ch}} \approx
12~\biggl(\frac{Z}{A}\biggr)^2~ M_{\odot}\label{s8}
\end{eqnarray}

and the equilibrium radius

\begin{eqnarray}
R_{\star}=
\frac{3^3}{2^4}\biggl(\frac{10}{\pi}\biggr)^{1/6}\frac{ea_0}{G^{1/2}m_p
\alpha^{1/2}}\frac{1}{AZ}\approx \nonumber\\
\approx 1.6\frac{R_{\odot}}{AZ} ,\label{s10}
\end{eqnarray}

where $M_{Ch}=\biggl(\frac{\hbar c}{G
m_p^2}\biggr)^{3/2}m_p=3.42\cdot 10^{33} g$ is the Chandrasechar's
mass, $M_{\odot}$ and $R_{\odot}$ are the mass and the radius of
the Sun, $m_p$ is the proton mass, $G$ is the gravitational
constant, $\alpha$ is the fine structure constant, $a_0$ is the
Bohr radius.

As the equilibrium number of nuclei inside a star is

\begin{equation}
N_{\star}=M_{\star}/A~m_p,
\end{equation}

with account of Eq.(\ref{s8}) and Eq.(\ref{s10}),

\begin{equation}
E_z \approx -0.8~\biggl(\frac{\hbar c}{a_0}\biggr)~Z^3 \approx
-5\cdot10^{-9}~Z^3~erg.\label{en1}
\end{equation}

Obviously, that  the heavy nuclei fusion is really accompanied by
a decreasing of electrostatic energy of plasma.

This energy, like the fusion energy, causes a calorification of a
star and can be radiated. This energy is related to the process of
fusion but it is of quite a different nature - this is the
gravitational energy which is caused by a changing of the
electrostatic energy of plasma.

\begin{figure}
\begin{center}
\includegraphics[5cm,1cm][20cm,11cm]{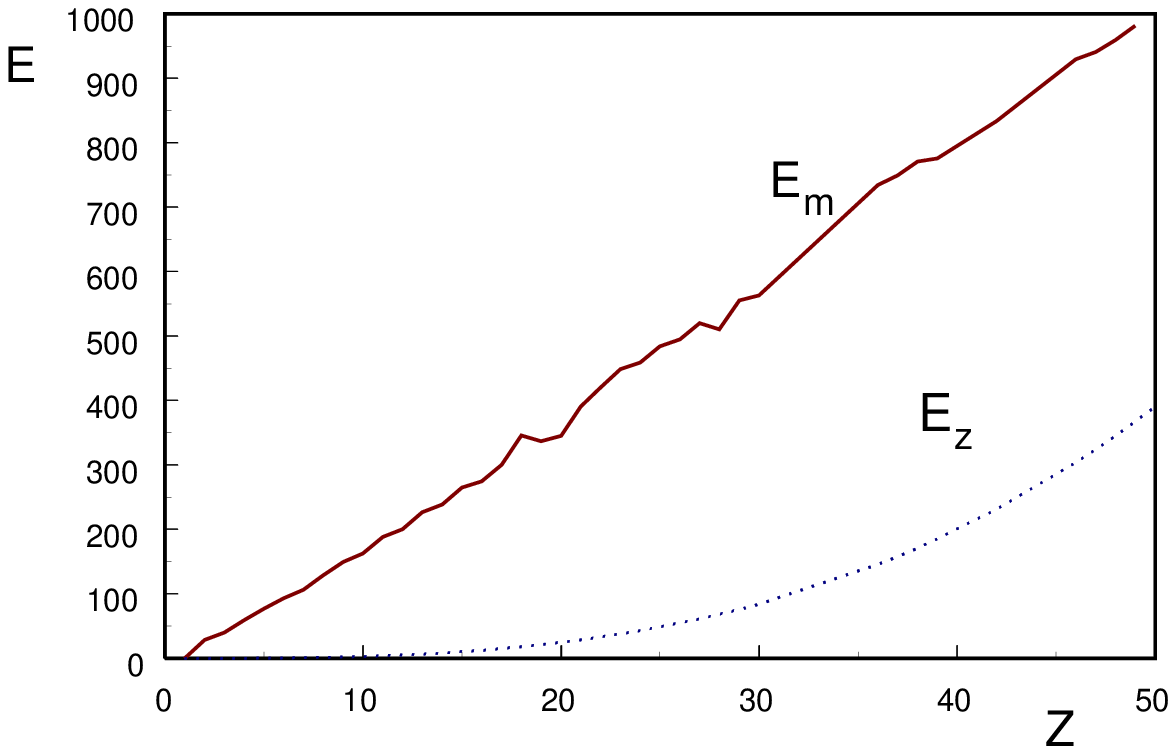}
\caption{ The plasma energy (in $Mev$) vs an averaged charge of
nuclei.} \label{dm}
\end{center}
\end{figure}

The dependencies of the transmutation energy and the electrostatic
energy of plasma on the averaged nuclear charge Z are shown in
Fig.{\ref{dm}}.

The luminosity of a star depends on the energy generation velocity
inside it:

\begin{equation}
\frac{dE}{dt} = \frac{dE_{m}}{dt}+\frac{dE_Z}{dt} = -(3\cdot
10^{-5}+1.5\cdot 10^{-8}Z^2)\frac{d Z}{dt} ~~\frac{erg}{s}.
\end{equation}

\begin{figure}
\begin{center}
\includegraphics[5cm,1cm][20cm,11cm]{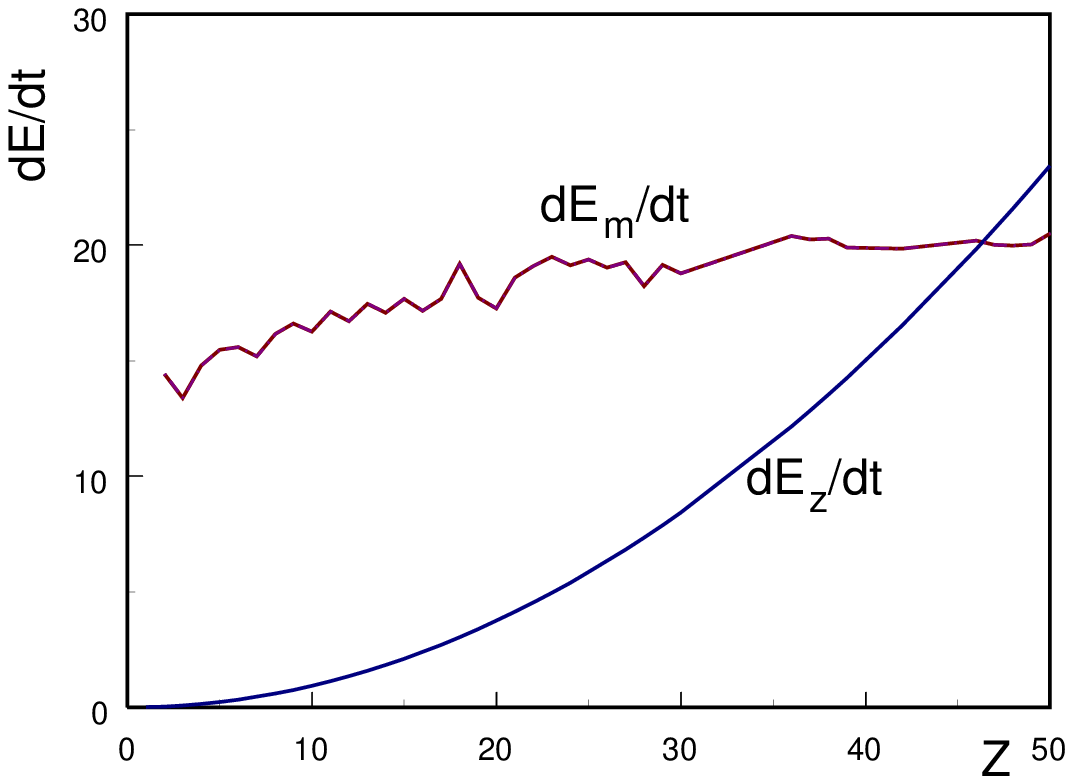}
\caption{ Velocity of energy generation (in $Mev\cdot dZ/dt$)
inside a star.} \label{ddm}
\end{center}
\end{figure}

The dependencies of energy generation velocities on Z are shown in
Fig.{\ref{ddm}}.

Essentially, the full radiation from the surface of a star depends
on both - the fusion energy and the decrease in a plasma
electrostatic  energy. Simultaneously, neutrinos are generated by
fusion processes only. It is known, that the neutrino flux (with
account of all fusion processes) is approximately two times less
than the value which can be obtain on the basis of full radiancy
of the Sun.

It is possible to conclude from Fig.{\ref{ddm}}, that an averaged
charge of nuclei of the plasma core in this case is close to
$Z\approx 45$.

\end{document}